\documentclass[conference]{IEEEtran}
\IEEEoverridecommandlockouts
\usepackage{cite}
\usepackage{amsmath,amssymb,amsfonts}
\usepackage{algorithmic}
\usepackage{graphicx}
\usepackage{textcomp}
\usepackage{xcolor}
\usepackage{url}
\def\BibTeX{{\rm B\kern-.05em{\sc i\kern-.025em b}\kern-.08em
    T\kern-.1667em\lower.7ex\hbox{E}\kern-.125emX}}
    
\usepackage{xspace}
\newcommand{\Abacus}[0]{\textsc{Abacus}\xspace}
    
\begin{document}

\title{Checkpointing with cp: the POSIX Shared Memory System}

\author{\IEEEauthorblockN{Lehman H.~Garrison}
\IEEEauthorblockA{\textit{Center for Computational Astrophysics} \\
\textit{Flatiron Institute}\\
New York, NY 10010 \\
lgarrison@flatironinstitute.org}
\and
\IEEEauthorblockN{Daniel J.~Eisenstein}
\IEEEauthorblockA{\textit{Center for Astrophysics} \\
\textit{Harvard \& Smithsonian}\\
Cambridge, MA 02138 \\
deisenstein@cfa.harvard.edu}
\and
\IEEEauthorblockN{Nina A.~Maksimova}
\IEEEauthorblockA{\textit{Center for Astrophysics} \\
\textit{Harvard \& Smithsonian}\\
Cambridge, MA 02138 \\
nina.maksimova@cfa.harvard.edu}
}

\maketitle

\begin{abstract}
We present the checkpointing scheme of \textsc{Abacus}, an $N$-body simulation code that allocates all persistent state in POSIX shared memory, or ramdisk.  Checkpointing becomes as simple as copying files from ramdisk to external storage.  The main simulation executable is invoked once per time step, memory mapping the input state, computing the output state directly into ramdisk, and unmapping the input state.  The main executable remains unaware of the concept of checkpointing, with the top-level driver code launching a file-system copy between executable invocations when a checkpoint is needed.  Since the only information flow is through files on ramdisk, the checkpoint must be correct so long as the simulation is correct.  However, we find that with multi-GB of state, there is a significant overhead to unmapping the shared memory.  This can be partially mitigated with multithreading, but ultimately, we do not recommend shared memory for use with a large state.
\end{abstract}

\begin{IEEEkeywords}
simulation, checkpointing, shared memory, ramdisk
\end{IEEEkeywords}

\section{Out-of-core model}
\Abacus is a cosmological $N$-body solver that integrates particle trajectories under mutual self-gravity in 6D phase-space from the smooth, nearly-uniform condition of the early universe to the clustered richness of filaments, halos, and voids seen in galaxy surveys today.  The depth and breadth of upcoming galaxy surveys like the DESI, the Dark Energy Spectroscopic Intrument, demand simulations with hundreds of billions or trillions of particles \cite{desi}.

Particle-based simulation codes are often memory limited, since the natural dimensions for improvement are to simulate physics at smaller scales or in a larger domain---either way, using more particles.  The floating-point throughput provided by GPUs has made processing such increasing numbers of particles tractable; in many cases, the challenge is now efficiently storing the state, and on distributed memory systems, communicating state between nodes.

\Abacus is designed to support simulations whether they fit in memory or not.  In the latter case, also known as ``out of core'', the state is buffered on disk.  A side-effect of this model is that the state \textit{is} the checkpoint: the only data flow between simulation time steps is through files on disk.  This enforces the completeness of checkpoints.

When running on a distributed-memory system where the state \textit{does} fit in memory, we preserve this file-oriented model by using POSIX shared memory, or ramdisk\footnote{We are using the term ``ramdisk'' colloquially, since, in Linux kernel parlance, a ramdisk is ``raw'' block device on top of which a file system may be created. We are using the related, but distinct, kernel tmpfs, which is more widely available.} \cite{shm_overview,tmpfs}.  The ramdisk is exposed as a directory, so checkpointing consists of launching a file-system copy in between time steps.  The simulation ``reads'' the shared memory with a memory map, thus avoiding a copy that a file-system read from ramdisk would incur.  However, this mapping procedure has its own overheads, which we will discuss below.

\section{Simulation flow}
\Abacus is divided into a top-level driver code and a simulation executable.  The driver code calls the executable in a loop, once per time step.  The task of the executable is to load the state at time $t$, called the \textit{read state}, compute forces on particles, update their kinematics, and write a new state at time $t+1$, called the \textit{write state}.  The executable is idempotent, relying on the top-level driver code to rename the write state to the read state between invocations.

The fact that a new executable is invoked for each time step, loading the state anew, ensures that the state \textit{is} the checkpoint.  There can be no side-channel information that flows through out-of-band allocations; any information that persists across time steps must be part of the state files.  This model is a strong enforcement of the completeness and correctness of checkpointing.  Furthermore, the executable can remain oblivious to the concept of checkpointing, leaving the driver code to handle the file-oriented tasks to which it is well-suited.

The state files themselves are raw binary representations of the phase-space particle information, with files divided into planar ``slabs'' of the simulation volume.  Within a slab, particles are ordered by cell (the atomic unit of our domain decomposition).  These files mirror the memory model used by the simulation.  Metadata is stored in a separate ASCII file in the state directory.

Each slab may have multiple \textit{slab types}, each stored in a separate file and representing one field, such as the positions, velocities, and particle flags.  When a given slab is requested by the code, the type is specified as well.  The request is processed by the \textit{slab buffer}, which computes the file path and determines whether it resides on ramdisk.  If so, the slab buffer instructs the \textit{arena allocator} to map the slab directly.  If not, the arena allocator makes a new allocation, and a file I/O request is passed to the I/O subsystem to read into that allocation.


The determination of whether a path resides on the ramdisk is done by string comparison of the path prefix.  Other, more robust methods were deemed either too complex or too expensive.


\section{Shared Memory}
POSIX shared memory is exposed on Linux via a tmpfs file system.  By default, it is mounted at \texttt{/dev/shm/} and has a capacity of half of the system's physical memory.  Files created in that directory have ``kernel persistence'', meaning they stay in memory until the kernel is terminated.

One way to use this ramdisk is as if it were an ordinary storage device, reading and writing with standard file I/O interfaces.  This will speed up I/O in most cases, but it will consume extra memory and the I/O will only be as fast as a memory copy.  This can be slow for large allocations---especially if the I/O is only using a single thread on a system with multiple sockets---and exerts unnecessary memory bandwidth pressure.

We instead opt to memory map the ramdisk files.  This can be thought of as getting a pointer directly into shared memory, avoiding any memory movement.  This is accomplished by getting a file descriptor with \texttt{open()}, setting the size with \texttt{ftruncate()} (if writing), and mapping the shared pages into user space with \texttt{mmap()}.

This model has been very successful in our code, with state files naturally serving as the checkpoint, and the actual backup to disk being as simple as launching a file system copy on each node.  We have confirmed that even though the shared memory is held by the kernel, the underlying pages obey user-space first-touch NUMA semantics.

The default ramdisk size limit on Linux systems is half of the system's memory.  This is not a limitation in our case, as roughly half of \Abacus's memory allocations are transient, mostly from kinematic data like particle accelerations.

\section{Deployment on Summit}
We ran a suite of simulations on the Summit\footnote{\url{https://www.olcf.ornl.gov/summit/}} supercomputer at Oak Ridge National Lab using this shared-memory checkpointing model.  Overall, it was very successful, with timed checkpoints running every few hours, and conditional checkpoints running before time steps that included on-the-fly analysis.  These time steps were considered ``riskier'' as they increased the memory footprint and the code path was dependent on the physical state of the simulation, increasing the chance of exposing a rare, corner-case bug.  The state copy from the nodes to the Alpine network file system took on average 2 minutes for 13 TB (6800 files) spread across 63 nodes, or about 1.7 GB/s/node.

The primary checkpointing failures were (i) a string of network failures triggered by the copy operation on multi-GB files, (ii) timeouts in the checkpointing caused by variable network file system performance, and (iii) user error in deleting the original checkpoint instead of the partial checkpoint in the case of checkpoint failures.


\section{Overheads}
\begin{figure}
    \centering
    \includegraphics{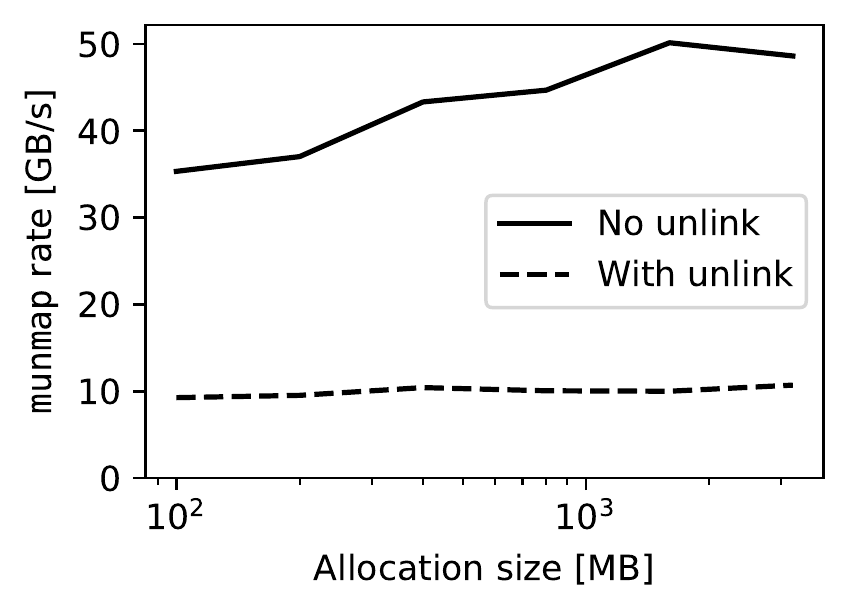}
    \caption{In the POSIX shared memory checkpoint model, all persistent allocations are made with \texttt{mmap()} and freed with \texttt{munmap()}. \texttt{mmap()} is fast, but \texttt{munmap()} is noticeably slow, especially when the corresponding file handle has already been unlinked or deleted (dashed line).  With checkpoints of 10s of GB, an unmap rate of 10 GB/s can be a bottleneck on the simulation performance.}
    \label{fig:overhead}
\end{figure}

We find that unmapping shared memory has a noticeable overhead that scales with the size of the mapping.  This is shown for a Linux Intel Skylake platform (page size 4096 bytes) in Figure~\ref{fig:overhead}.  All mappings and unmapping were performed with a thread affinity fixed and on a single NUMA node.  Two cases are shown: with and without the underlying \texttt{/dev/shm/} file name being unlinked (deleted) before the unmapping.  If the file name has not been unlinked, then the unmapping is fast (10s of GB/s).  If the name has been unlinked, then the unmap runs at 10 GB/s independent of the size of the mapping.  This rate is similar to the \texttt{memset()} speed and suggests some kind of operation (zeroing?) is occurring on the contents of the pages, not just the page tables.  This work can be assigned to its own thread, but in simulations, we have observed performance degradation in other areas of the code while \texttt{munmap()} is running in a separate thread.  Memory bandwidth pressure may be to blame.

We confirm that an ordinary \texttt{malloc()}/\texttt{free()} pair does not exhibit this behavior.  The Summit platform exhibited this same pattern of overheads, despite being a IBM POWER9 platform with 64 KB pages.

For certain allocations (write state slabs), we can skip the \texttt{munmap()} call, as it will not free any memory, because the underlying file handle must persist until the next time step.  However, we find that doing so simply defers the unmapping cost to the program termination.  Similarly, performing an unlink after the unmapping, rather than before, just shifts the time differential into the unlink.

These overheads may be similar or even smaller than methods that stage checkpoints in a main memory buffer or a burst buffer---a write to a burst buffer will likely be slower than 10 GB/s.  However, the overheads are incurred for every time step (typically $\mathcal{O}(1000)$), rather than a few times per simulation.  A hybrid method that allows the simulation executable to run multiple time steps in memory then switch to the ramdisk method just for the checkpoint step may be superior, at the cost of increased code complexity.

We surmise that the shared memory system was designed to facilitate lightweight inter-process communication, and not allocations of dozens of GB.  Because our code requires a large amount of state relative to the time it takes to process it, it is suboptimal to use POSIX shared memory as the only way to pass information between time steps.  However, the correctness enforced by the out-of-core model is a useful property. This model may be appropriate for codes with smaller state or a higher compute density (ratio of compute work to state size).

\section*{Acknowledgment}
The authors would like to thank the co-developers of the \Abacus code: Marc Metchnik, Doug Ferrer, and Phil Pinto.  Abacus development has been supported by NSF AST-1313285 and more recently by DOE-SC0013718, as well as by Simons Foundation funds and Harvard University startup funds.  NM was supported as a NSF Graduate Research Fellow. The Summit simulations have been supported by OLCF projects AST135 and AST145, the latter through the Department of Energy ALCC program.

\bibliographystyle{./bibliography/IEEEtran}
\bibliography{./bibliography/IEEEabrv,./bibliography/biblio.bib}

\end{document}